# Securing Safety in Collaborative Cyber-Physical Systems through Fault Criticality Analysis

Manzoor Hussain# • Nazakat Ali#• Jang-Eui Hong†

## ABSTRACT

Collaborative Cyber-Physical Systems (CCPS) are systems that contain tightly coupled physical and cyber components, massively interconnected subsystems, and collaborate to achieve a common goal. The safety of a single Cyber-Physical System (CPS) can be achieved by following the safety standards such as ISO 26262 and IEC 61508 or by applying hazard analysis techniques. However, due to the complex, highly interconnected, heterogeneous, and collaborative nature of CCPS, a fault in one CPS's components can trigger many other faults in other collaborating CPSs. Therefore, a safety assurance technique based on fault criticality analysis would require to ensure safety in CCPS. This paper presents a Fault Criticality Matrix (FCM) implemented in our tool called CPSTracer, which contains several data such as identified fault, fault criticality, safety guard, etc. The proposed FCM is based on composite hazard analysis and content-based relationships among the hazard analysis artifacts, and ensures that the safety guard controls the identified faults at design time; thus, we can effectively manage and control the fault at the design phase to ensure the safe development of CPSs. To validate our approach, we introduce a case study on the Platooning system (a collaborative CPS). We perform the criticality analysis of the Platooning system using FCM in our developed tool. After the detailed fault criticality analysis, we investigate the results to check the appropriateness and effectiveness with two research questions. Also, by performing simulation for the Platooning, we showed that the rate of collision of the Platooning system without using FCM was quite high as compared to the rate of collisions of the system after analyzing the fault criticality using FCM.

# 협업 사이버물리시스템의 결함 치명도 분석을 통한 안전성 확보

후세인 만즈르# • 나자캇 알리# • 홍장의†

## 요 약

협업 사이버-물리 시스템(Collaborative Cyber-Physical Systems, CCPS)은 물리 세계와 사이버 세계가 밀접하게 결합하여 공동의 목표를 달성하기 위하여 협업을 수행하는 시스템이다. 한편, 단일 사이버-물리 시스템(Cyber-Physical System)의 경우에는 ISO 26262 또는 IEC 61508과 같은 표준을 따르거나 다양한 위험 분석 기법을 적용함으로써 그 안전을 확보할 수 있다. 그러나 CCPS에서는 협업을 수행중인 한 CPS의 결함으로 인하여 다른 협업 중인 CPS에게 수많은 결함을 발생시키기 때문에 안전의 확보가 매우 어렵다. 본 논문에서는 이러한 CCPS의 위험을 분석하여 안전을 확보하기 위해 복합적인 위험 분석과 위험 분석 산출물 사이의 관계를 기반으로 하는 위험 치명도 매트릭스(Fault Criticality Matrix, FCM)를 제시한다. FCM에서는 결함, 결함의 치명도, 안전 가드와 안전 가드의 발생 확률, 결함의 영향 및 순위를 나열하여 분석한다. 안전 엔지니어는 이를 통해 시스템의 설계 단계에서 각 결함의 치명도와 영향을 분석하고, 설계된 안전 가드를 통해 식별된 고장을 효과적으로 관리하고 제어함으로써 안전한 CPS를 개발할 수 있다. 제시된 방법의 유용성을 확인하기 위해 CCPS의 대표적 예인 군집주행에 대하여 사례 연구를 수행하였다. 본 연구에서 개발된 도구를 사용하여 군집주행 시스템에 FCM을 적용함으로써 상세한 결함 치명도 분석을 수행하였고, 분석 결과는 적합성과 효과성 관점에서 점검되었다. 또한 군집 주행에 대한 시뮬레이션 수행을 통해 FCM을 사용하여 결함 치명도를 분석한 군집주행 시스템이 발견된 모든 결함을 완화시켜 충돌 가능성을 크게 낮추었음을 보였다.

## 1. Introduction

† 정 회 원 : 충북대학교 소프트웨어학과 교수
# 학생회원 : 충북대학교 컴퓨터과학과 석박사과정

Collaborative CPSs are tightly coupled with cyber and physical components, massively interconnected, and collaborate by sharing information and resources to achieve common goals. The tight coupling of cyber and physical components has many advantages, but it also creates new challenges like safety issues that have not been considered in the traditional computing domain. The development of smart and context-aware mission-critical systems has drastically risen in recent years. Power grids, water distribution networks, smarts factory, medical CPS, autonomous platoon driving systems are examples of complex cyber-physical systems. The safety assurance of these safety-critical systems becomes very important because any mishap in these systems can cause property loss, environmental damage, and even loss of human lives. As the collaborative CPS's components or subsystems are massively interconnected, their collaboration creates safety concerns because a minor fault in one system can activate many other faults in other collaborating systems.

In our previous work [1], we worked on fault traceability in collaborative CPSs, and we presented a Fault Traceability Graph (FTG) to visualize the impact of faults and their origin. The fault propagation graph also visualizes the fault route as a means of traceability. This traceability graph was based on content relationships among the hazard analysis artifacts. A Fault Traceability Matrix (FTM) was also introduced as a means of another representation of traceability of faults in CPSs. However, in the FTM, fault ranking and the effect of safety guards on a particular fault were not well presented. The criticality calculation of each fault and its rankings before and after supplying corrective actions was not considered. The probability of occurrence of faults and safety guard is the building block of our approach. In previous work [1], we did not present the probability of the occurrence of faults and safety guards. As collaborative CPSs are safety-critical and mission-critical systems, fault traceability is very important to determine the fault routes, their origin and to determine that the system must meet the safety goals, and all identified hazards were eliminated [2] or decreased their criticality to an acceptable level. In [3, 4], the term traceability is defined by investigating the links between hazard analysis artifacts and design documents. However, the criticality analysis of faults in CPSs was not investigated. The propagation of the fault in collaborative CPSs is very critical for mission-critical systems. In a safety-critical system, even a minor fault cannot afford as it can activate many other faults in the collaborating systems. Therefore, a criticality analysis technique would be required to determine how much a specific fault is a threat to the safety of collaborative CPSs.

Hazard analysis helps safety engineers to identify faults and to resolve the insufficiency of safety-related functions, and some hazard analysis provides information about safety guards as mentioned in [5]. Fault Tree Analysis (FTA), Failure Mode Effect Analysis (FMEA), and Event Tree Analysis (ETA) are the most fundamental hazard analysis techniques used to perform the hazard analysis. The main objective of FTA is to determine the root causes of the specified fault. While on the other hand, the ETA is used to determine whether the safety guards sufficiently control an undesired event or not. However, the FMEA is used to evaluate the failure mode and effect of components, assemblies, and subsystems.

## 1.1 Motivation

The collaborative nature of CPSs creates safety related challenges as CPSs work in the physical environment. Hence, fault reduction and prevention approaches must be considered at design time. As collaborative CPSs are very complex and heterogeneous, a single hazard analysis technique is insufficient to guarantee safety for such kinds of collaborative CPSs. Therefore, we proposed a composite hazard analysis technique in our previous work [6] based on content relationships among the hazard analysis techniques such as ETA, FMEA, and FTA. The composite hazard analysis technique can be used to analyze the hazard for complex systems like collaborative CPSs. Due to these properties such as massively collaboration of cyber-physical systems, fault criticality assessment is a challenging task. We assume that the proposed fault criticality matrix can enable the safety engineers to evaluate faults' criticality and their impact on other collaborating systems. It also enables the safety engineers to manage the fault and its propagation scope. The proposed Fault Criticality Matrix (FCM) ensures that the safety guard can sufficiently control the specified fault. As it is a basic principle of the cyber-physical system, all potential faults that may lead to undesired events must be illuminated to ensure the developed system is safe enough from harmful to humans and its environment.

Intending to present a criticality analysis method to determine the fault criticality in collaborative cy-ber-physical systems, we first perform the composite hazard analysis of collaborative CPSs by using our developed tool called CPSTracer. Then we generate the FCM for CCPS. By extending our previous work [6] in which we presented only a fault traceability graph, now present the fault criticality matrix to perform criticality analysis of faults in collaborative CPSs.

### 1.2 Contributions

In the paper, we make the following contributions:

- First, we propose FCM to analyze the criticality of faults that can occur during CPS collaboration. The proposed FCM can demonstrate the criticality of faults with their ranking. It also shows the criticality of faults before and after supplying safety guards. This clear representation of criticality analysis helps to ensure that the supplied safety guards are enough or not.
- The proposed FCM was implemented and integrated with our tool CPSTracer. So, the information about fault criticality can support safety engineers to perform the safety analysis process seamlessly.
- We present a case study on the Platoon driving system (a collaborative CPSs) and generate the FCM to perform the criticality analysis of faults associated with the platooning system.

## 2. Related Work

The authors in [7] proposed a framework called SafeTrace that can manage the traceability among the safety requirements, design, and safety analysis artifacts in the medical device. The authors explored the links between the hazard analysis artifacts, requirement artifacts, and design artifacts to see whether the change of requirement or design may affect the safety in the system. More specifically, the authors present the trace links between the design documents and basic events in FTA and between the top event in FTA and the requirement document. However, the criticality analysis of hazard artifacts was not determined in this approach.

In an another study [8], the authors proposed an approach to develop safe and secure adaptive collaborative systems with runtime safety guarantees. The aim of this study was to develop run-time behavioral models for collaborative distributed systems and hazard analysis techniques to support safety and ensuring cybersecurity for assumptions made in the model. In this case, it is needed to design the behavioral models, and techniques to analyze and check the safety and cyber-security both at design time and runtime. In this approach, the authors considered the platooning system as a case study to validate their approach. However, the fault impact analysis and analysis of criticality of faults were not considered. It also does not support composite hazard analysis as a complex collaborative system's safety can not be guaranteed by a single hazard analysis technique.

Medawar et al. [9] proposed an approach that provides safety in the platooning CPSs within the safeCOP project. The aim of safeCOP project was to analyze safety-critical systems that need to provide safety assurance. The authors further argued that safety in cooperative CPSs is a challenging task because safety can only be ensured if all the systems cooperate safely during their operations. The authors used a runtime manager to handle continuous safety assurance for a truck platooning use case. The runtime manager's functionality can be fully utilized when the safety contracts for safety analysis of the local system and the overall cooperative system are specified. However, the authors did not consider composite hazard analysis, the fault criticality, and impact analysis in safeCOP project.

Kim et al. [10] developed a tool called NuDE 2.0 for safety analysis and verification of safety-critical systems. However, NuDE 2.0 does not support collaboration among multiple CPSs.

The authors in [11] proposed an architecture-based approach to investigate the human-robot collaboration and its capabilities to ensure human safety in a production environment. The authors considered a shared fenceless working space where robots, humans, and other moving objects may operate. The aim of this study was to explore the response time needed to detect the human and activate the safety signals to ensure human safety. The shorter the detection time, the more rapid collision avoidance strategies opted. On the other hand, the longer the response time and activating safety signal, the more chances to occur potential hazards.

The authors in [12] proposed an approach to address the fault criticality and interaction in complex consumer communications based on FTA. In this study, the authors outlined the limitation of traditional FTA. The fields and manufacturing failure modes, mechanism, and their interactions were arranged into a unified rank system to determine the criticality. This study also presents a new failure tree to demonstrate the criticality of faults.

## 3. Proposed Approach: Collaborative Behavior and Composite Hazard Analysis of CPSs

Collaborative CPSs are those systems that perform individual tasks to achieve a common goal. Due to the massively interconnected and heterogeneous subsystem, their collaboration creates safety concerns. However, hazard analysis in CPSs makes it possible for safety engineers to identify the potential hazards and provide the potential safety guards to mitigate the faults. Therefore, we proposed an approach to analyzing the safe operability of the collaborative system by supporting the composite hazard analysis for CPSs as a hazard analysis for single system does not guarantee safety for collaborative systems. Figure 1. illustrates our proposed approach.

In our proposed approach, we first analyze the collaborative nature of the cyber-physical system. As in collaborative CPSs, a fault in one CPS can activate other faults in other CPSs to whom it may collaborate. It means that a fault in, if occurred, in one system can propagates to other collaborating CPSs. As a result, the collaborative CPSs fail to achieve their final goal. Therefore, based on our previous work [1], we used our defined content relationships among the hazard analysis techniques such as FTA, FMEA, and ETA to perform composite hazard analysis for collaborative CPSs. Another reason for this relationship is to envision the relationship among faults coming from different collaborating systems and the impact of specific faults on other systems. As we mentioned earlier that the failure of one collaborative CPS may affect other collaborating CPS. Therefore, to analyze the fault propagation and its routes, we introduced FTG in our previous work. However, the criticality analysis of faults was not considered at that time. Now, extending our previous work, we present a criticality matrix to perform criticality analysis in collaborative CPS. The proposed FCM ranks the criticality of faults based on the probability of occurrence of a fault and its impact on other faults. This means that how many other faults are influenced by a particular fault. There are two rankings in the FCM matrix, in the first ranking criteria of faults, the safety guards are not considered to show the actual criticality of fault without providing a safety guard. After applying the safety guards to the faults, we rank the fault again to check whether the supplied safety guards were enough to cope with the specified fault or not, as well as to check the effect of the safety guard on fault criticality. Definitely, after supplying some safety guards, the criticality of faults is reduced.

### 3.1 Collaborative Behavior of CPS

To analyze the collaborative behavior of CPSs, we use the platooning system as our case study to justify our approach. The platoon driving system forms a platoon of several vehicles maintaining a short inter-vehicle distance with the preceding vehicles to improve traffic flow, reducing traffic congestion, and reducing fuel consumption [13]. Figure 2 shows an example of the platooning system.

In the platooning system, the head of the platoon is called the leader, and the following cars behind the leader are called followers. Cooperative Adaptive Cruise Control (CACC) is used vehicle to vehicle communicate in the platooning system. Each vehicle in the platooning system is independent CPS and can use the Adaptive Cruise Control (ACC) unit when necessary. As in the platooning system, the distance among the vehicles is supposed to be very short; therefore, a fault in the leader vehicle can be propagated to the other vehicles, as a result, collision among the vehicles may occur. For example, if the leader vehicle sends a wrong command of acceleration to follower vehicles, then the collision will potentially occur. Even though a platooning system developed with safety requirements, it is not possible to consider all possible failure conditions in runtime.

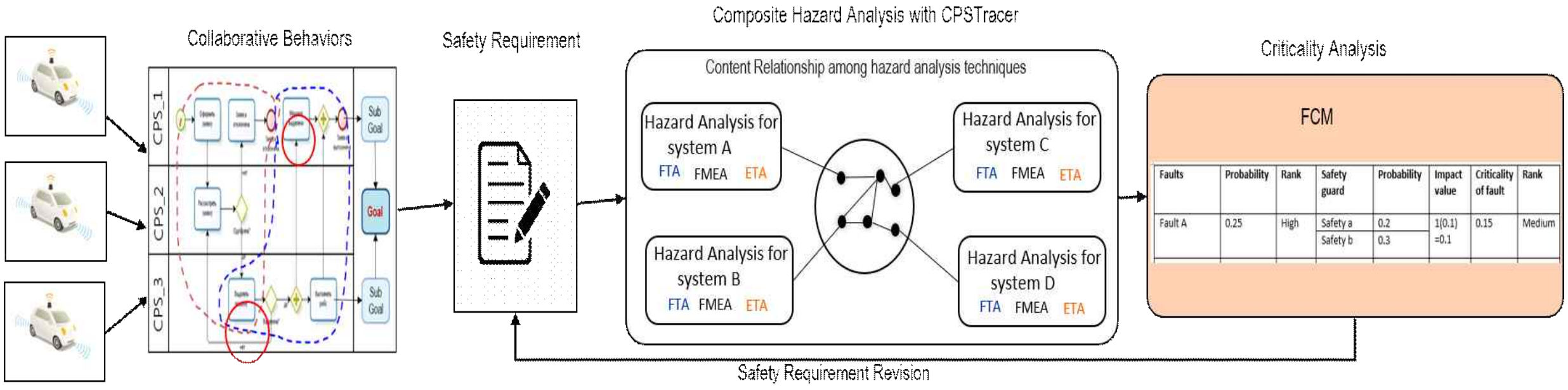

Figure 1. A Proposed Approach for Safety Assurance through Fault Criticality Analysis in Collaborative CPS.

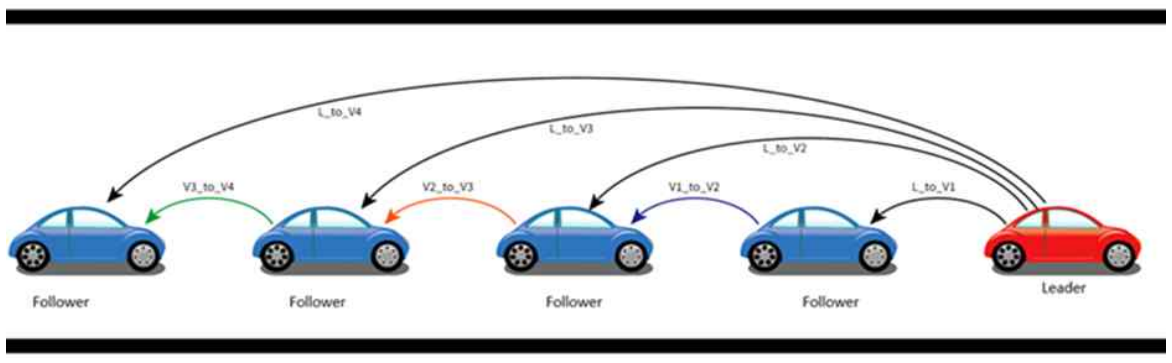

Figure 2. Example of a Platooning System.

### 3.2 Safety Requirements

The safety requirements are defined to ensure the safety in systems. These requirements are first specified at a high level, refined, and then supplied to the designer. Safety requirements are generated after analyzing the collaborative behavior of CPSs in our approach to performing composite hazard analysis on specified safety requirements. After extracting the safety requirements from the collaborative behavioral analysis of CPSs, we supply the safety requirement as input to our composite hazard analysis technique to perform hazard analysis. In the next step, we perform the criticality analysis of faults to verify whether identified faults are removed from the system or not. This process is a close loop, and it continues until all potential faults are mitigated from the system or an acceptable level of safety is achieved, and the safety requirements are also revised according to the status of the fault.

### 3.3 Composite Hazard Analysis of the Platooning System

Usually, in the platooning system, the vehicles form a group and collaborate to reduce traffic congestion by reducing the inter-vehicle distance. However, the shorter distance among the participating vehicles creates safety concerns for vehicles in the platooning system. We assume that the safety of CCPS can be ensured by performing the hazard analysis to explore potential faults in the system. For this purpose, we used our composite hazard analysis tool called CPSTracer to perform hazard analysis. The composite hazard analysis technique is based on four kinds of content relationships, influence, inheritance, supplement, and overlap relationships among the hazard analysis techniques such as FTA, FMEA, and ETA [1]. The definitions of relationships are as follows:

1) Influence Relationship

The influence relationship exists among the faults in participating CPSs in which a fault in one participating system causes the activation of another fault in CCPSs.

2) Inheritance Relationship

This relationship is established when two or more participating systems in CCPS share the same operational and functional constraints. The inheritance relationship also exists among the faults of the participating systems.

3) Overlap Relationship

This relationship exists when the consequences of the failure of one system are the same as the failure consequences of another system. This relationship exists among the outcomes/consequences of the failure of systems in collaborative CPSs.

4) Supplement Relationship

This relationship exists among the safety guards and

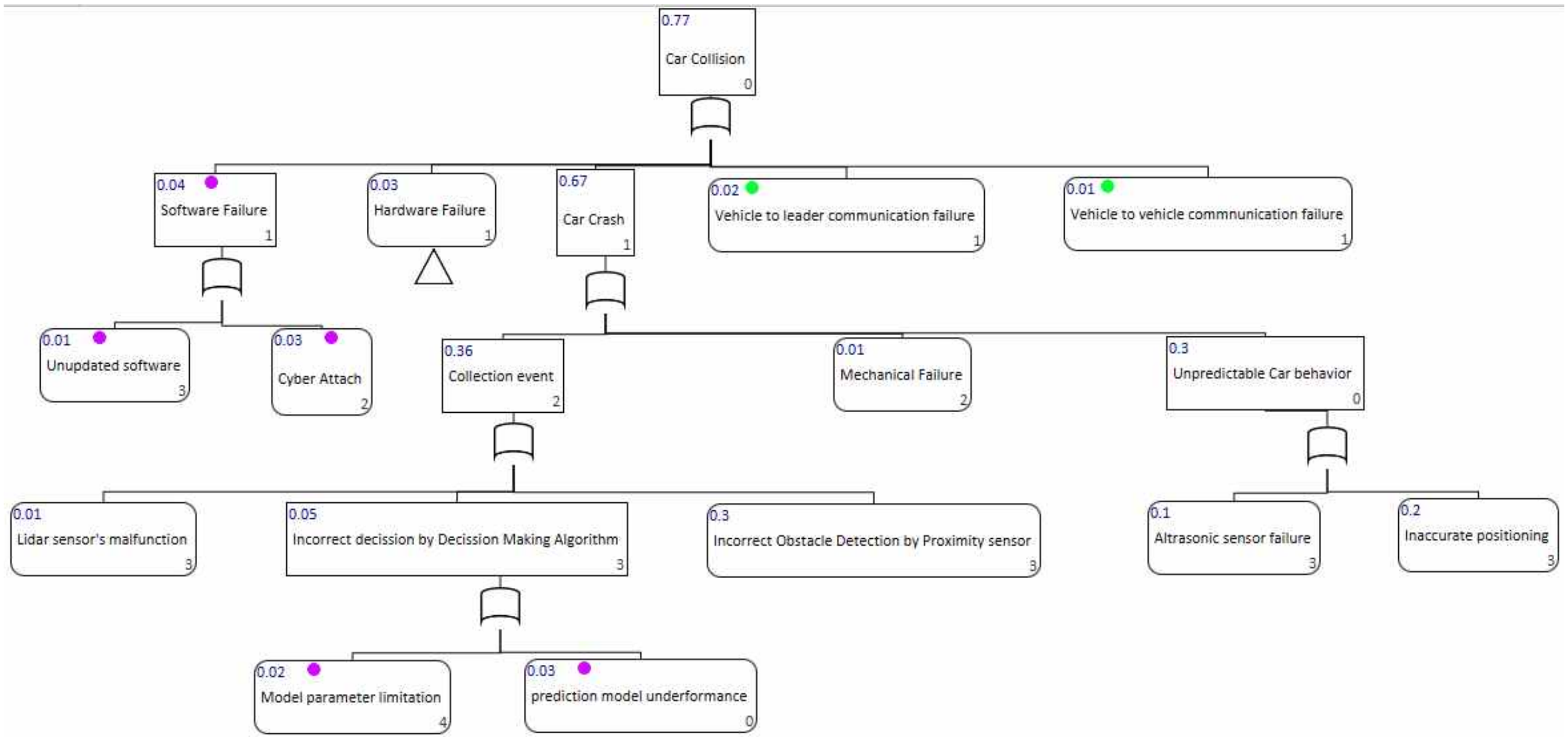

Figure 3. Fault Tree Analysis of the Platooning System.

FMEA

System: Autonomous Car Platooning Subsystem Mode/Phase

| Item | Failure Mode | Casual Factors | Immediate Effect | System Effect | Probability of Occurance | Safety Gaurd | Probability of Safety Ga... |
|---|---|---|---|---|---|---|---|
| Communication Module | Vehicle to vehicle Com... | Communicational infrast... | Unable to communicate ... | Communicational Failure | 0.1 | Reduce Speed and exit p... | 0.32 |
| Camera Sensor | Detection Failure | Weather Condition | Fail to detect opstacles | Detection Sensor Failure | 0.02 | Exit Platooning and redu... | 0.12 |
| Object Localization Algor... | Wrong prediction of fron... | Limitation of Object Loca... | Wrong prediction of fron... | Fail to predict front car position and distance | 0.03 | Decrease Speed | 0.04 |
| Lidar Sensor Failure | Sensing Failure | Weather Condition | Fail to determine opstacles | Lidar Sensor Failure | 0.03 | Check for secondary sen... | 0.03 |
| GPU | Processing time exceeds | Low pocessing powered ... | Fail to process sensory d... | Processing function failure | 0.1 | Reduce speed to exit pla... | 0.5 |
| Proximity Sensor | Power Supply Failure | Shortcuit at power suply... | Fail to detect near by car | Proximity Sensor Failure | 0.03 | Reduce Speed | 0.05 |
| Decission Making Algorit... | Inappropriate Decission | Fail to recognize scenarion | Wrong Decission | Unpredictable Car Behav... | 0.01 | Turn of light and safely e... | 0.02 |

Figure 4. Failure Mode and Effect Analysis of the Platooning System.

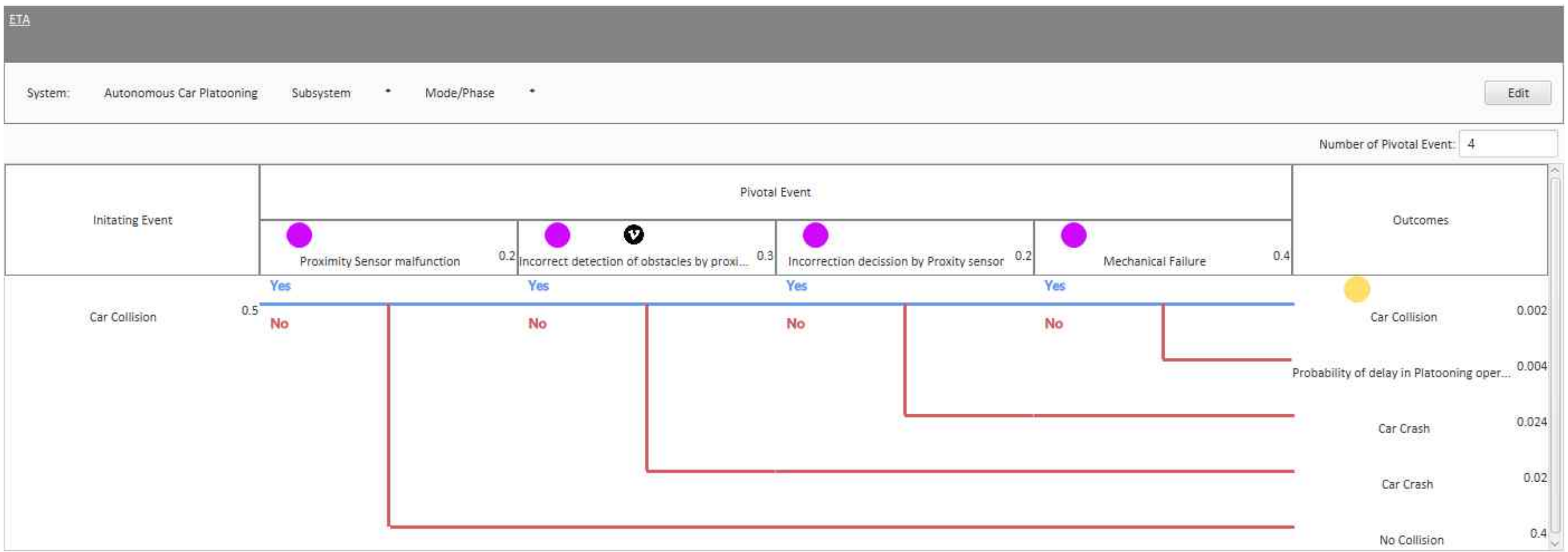

Figure 5. Event Tree Analysis of the Platooning System

faults in collaborative CPSs. When a system has a safety guard to cope with identical faults in another system, then this relationship is established between the safety guard and that particular fault. This means that the safety guard for a fault of a system can be supplied to another identical fault of a system in CCPs.

FTA is widely used for hazard and risk assessment in CPSs. In our approach, we first perform an FTA as part of composite hazard analysis to identify the root cause of failure of our case study i.e., the platooning system. We assume that one of the reasons for the platooning system failure is *Car Collision* (i.e., the top event in FTA). The top event in FTA is the failure of the system as a whole which means in the case of the platooning system, the participating vehicles do not collaborate as a result, the collision has happened. An FTA of five levels for the platooning system is constructed as shown in Figure 3. The intermediate and the basic events in FTA are root causes of the top event (i.e., *Car Collision*).

The FMEA is a structured method for system safety analysis to identify, evaluate, and score the potential failure for the system and its effect on the system. In our composite hazard analysis technique, we performed FMEA for the platooning system for its potential hazards. We perform FMEA on *Communication Module, Decision Making Algorithm, Object Localization Algorithm, Proximity Sensor, Camera Sensor, and Lidar Sensor*, etc. Figure 4 shows the FMEA for the platooning system performed in our composite hazard analysis tool. We introduce new columns in FMEA for safety guards and their probability of occurrence and probability of occurrence of failure modes.

In the last step of our composite hazard analysis technique, we perform event tree analysis for the platooning system. ETA shows all potential outcomes stemming from the undesired event and takes into additional events and factors i.e., whether or not the installed safety barriers are working. ETA is used to identify the possible potential accident scenario and sequences in the complex system. Due to its efficacy in analyzing hazards in complex systems, we assume that it is good for collaborative CPSs as well. Figure 5 shows the ETA for the platooning system.

## 4. Fault Criticality Analysis

The collaborative cyber-physical systems are massively interconnected and heterogeneous and collaborate with other CPSs to achieve complex tasks. Due to these properties of CCPS, more and more effective hazard analysis technique providing better traceability, impact analysis, and criticality analysis is required to mitigate all potential hazards from the systems. Determining fault criticality in the CPSs is a very challenging task especially in collaborative CPSs.

Therefore, a matrix called Fault Criticality Matrix is proposed to perform criticality analysis in CCPS with the following definition.

**[Definition]** Fault Criticality Matrix (FCM) is a tabular form that contains probability, criticality, and other data for a specific fault. The table consists with the tuple of <Fault, P. C, Rank, SG, P(SGs), IV, FC, $Rank_s$>, where each element is explained as follows and also shown in Table 1.

In the proposed FCM, we organize the identified faults in the first column i.e., the "*Fault*" column after performing the composite hazard analysis using our tool. In the second column "*P*", we arrange the probability of occurrence of each fault. In the third column ("*C*"), we calculate the criticality of fault without considering the safety guard to determine the actual fault criticality. The ranking of the faults is done based on the criticality of the fault which is organized in the column "*Rank*".

The safety guard column (i.e., "*SGs*") in our proposed FCM is consists of all safety guards provided to each fault and the probability of occurrence of safety guard is organized in column "*P(SGs)*". In column "*IV*", the impact value for each fault is placed. Impact value is the number of other potential faults that can be activated by a particular fault. This value is determined by the number of *influence, overlap,* and *inheritance* relationships that a fault makes with other faults. One influence relationship or one inheritance relationship or one overlap relationship with other faults have the impact value of 0.1.

For example, if a fault is making influence relationships with two other faults, then the impact value (*IV*) will be 0.2. Same as if a fault is making one inheritance relationship and two influence relationship, then the impact value (*IV*) will be 0.3. The second last column is "*FC*" which is the criticality of fault after supplying the safety guards. This column is introduced to show the effect of the safety guards on fault criticality. The "*Rank*" column ranks the criticality of faults after supplying a safety guard to show the effect of the safety guard on the criticality of each fault. Table 1 shows the template of the proposed fault criticality matrix. The ranking criteria for determining the criticality of faults are shown in Table 2.

Table 1. Template for Fault Criticality Matrix.

| Fault | P | C | Rank | SG | P(SGs) | IV | FC | $Rank_s$ |
|---|---|---|---|---|---|---|---|---|
| k | $X_k$ | $X_k$+N | High | i | $S_i$ | N | C | Medium |
| k | $X_k$ | $X_k$+N | Medium | i | $S_i$ | N | C | Low |
| k | $X_k$ | $X_k$+N | Low | i | $S_i$ | N | C | Negligible |

P: Probability, C: Criticality, SG: Safety Guard, P(SGs): Probability of Occurrence of Safety Guards, IV: Impact value FC: Fault Criticality (after safety Guard), $Rank_s$: Rank after the safety guard

### 4.1 Fault Criticality Calculation

The criticality of fault is represented by *FC*. Let us consider S ($s_1$, $s_2$, $s_3$, .., $s_i$) is the probability of safety guards where i=1, 2, 3... and X ($x_1$, $x_2$, $x_3$, ⋯, $x_k$) is the probability of faults where k=1, 2, 3,⋯ The safety guard $S_i$ is supplied to mitigate the impact of fault $X_k$ from the systems where i is the number of safety guards and k is the number of faults. The impact value of fault is represented by N. The value N is defined to be 0.1 if the fault has an impact on only one fault. If the faults impact multiple faults, then each impact would have a 0.1 value and the final impact value will be the sum of all values. The probability of occurrence of each fault, and safety guard is taken from the respective hazard analysis technique and the impact value is taken from the number of influence, overlap, and inheritance relationship that a fault is making with other faults. Now the value of fault criticality is determined by using equation (1).

$$FC = (X_k + N)\forall k - \sum_{i \in I} Si \qquad (1)$$

The minus or zero value of criticality of any fault indicates that the faults do not affect the system as the fault's criticality is mitigated by supplying the safety guards. If the criticality of fault still does not change after supplying the safety guard, then the supplied safety guard is not enough to mitigate the fault of the system. If the criticality of faults does not reduce, safety engineers must focus on such kinds of faults because these faults are more critical for system’ s safety. Reduction in criticality indicates that the criticality is reduced however, for the failure-free system, all potential faults must be mitigated or should reduce their criticality to zero. For this purpose, after applying the safety guard, safety verification must be done, and then revise the safety requirements. This process must continue until all safety goals are achieved.

Table 2. Ranking Criteria of the Criticality of faults.

| Fault Criticality Value (C) | Ranking |
|---|---|
| C <= 0.0 | No effect |
| 0.0 ~ 0.005 | Negligible |
| 0.005 ~ 0.01 | Low |
| 0.01 ~ 0.15 | Medium |
| 0.15 ~ 0.4 | High |
| 0.4 ~ 0.6 | Very High |
| 0.6 ~ 1.0 | Catastrophic |

### 4.2 Fault Criticality Analysis of the Platooning System

Cyber-physical systems are safety-critical. Renowned safety-critical CPSs, for example, self-driving cars, manufacturing plants, sensitive medical devices, and nuclear reactors, sometimes exhibit unsafe behaviors and which results in huge loss of properties and environmental damages. Addressing CPSs safety includes designing a safe cyber-physical system and verifying the CPS's safety until the desired safety goal is achieved. A term recently used for safe system designing is safety-guided designing which is used to designing the safe system and its safety verification. Hazard analysis is the crucial activity for safety-guided design which can be a challenging task for complex dynamics systems like collaborative CPSs. As the collaborative CPSs consist of complex, dynamic, and highly interconnected systems, determining the criticality of fault is also a challenging task. It is a very complicated task to determine which faults are more critical to systems safety and which have less criticality to breach the system safety. Therefore, we present the FCM to perform the criticality analysis of fault. This matrix is updated until the criticality of all faults is reduced to zero and the safety goals are achieved.

The platooning driving system is a collaborative CPS in which the leader and follower vehicles collaborate to reduce traffic congestion by maintaining a short inter-vehicle distance. However, the short distance between participating vehicles, relying on sensors creates safety concerns such that any fault in the participant vehicle can cause the whole platooning system collision. After analyzing the platooning system with our composite hazard analysis tool, we established content relationships among the hazard analysis (i.e., FTA, FMEA, and ETA) artifacts using our previously defined relationship [1], and we generated the Fault Criticality Matrix to perform the criticality analysis of potential faults for the platooning systems as shown in Figure 6.

Let us take the example of potential fault *Detection Failure.[Autonomous Car Platooning.FMEA_0]* which is discovered using FMEA as part of our composite hazard analysis technique. Its probability is 0.002 and criticality is 0.32 and it has an impact value of 0.3 which means that it is influencing three other faults in the platooning system. After calculating its criticality and before applying the safety guard, we can see that the criticality is "*High*" in the "*Rank*" column. However, after applying the safety guard "*Reduce Speed and exit platooning.[Autonomous Car Platooning.[FMEA_0]]*", we calculated the fault criticality again and the crit-

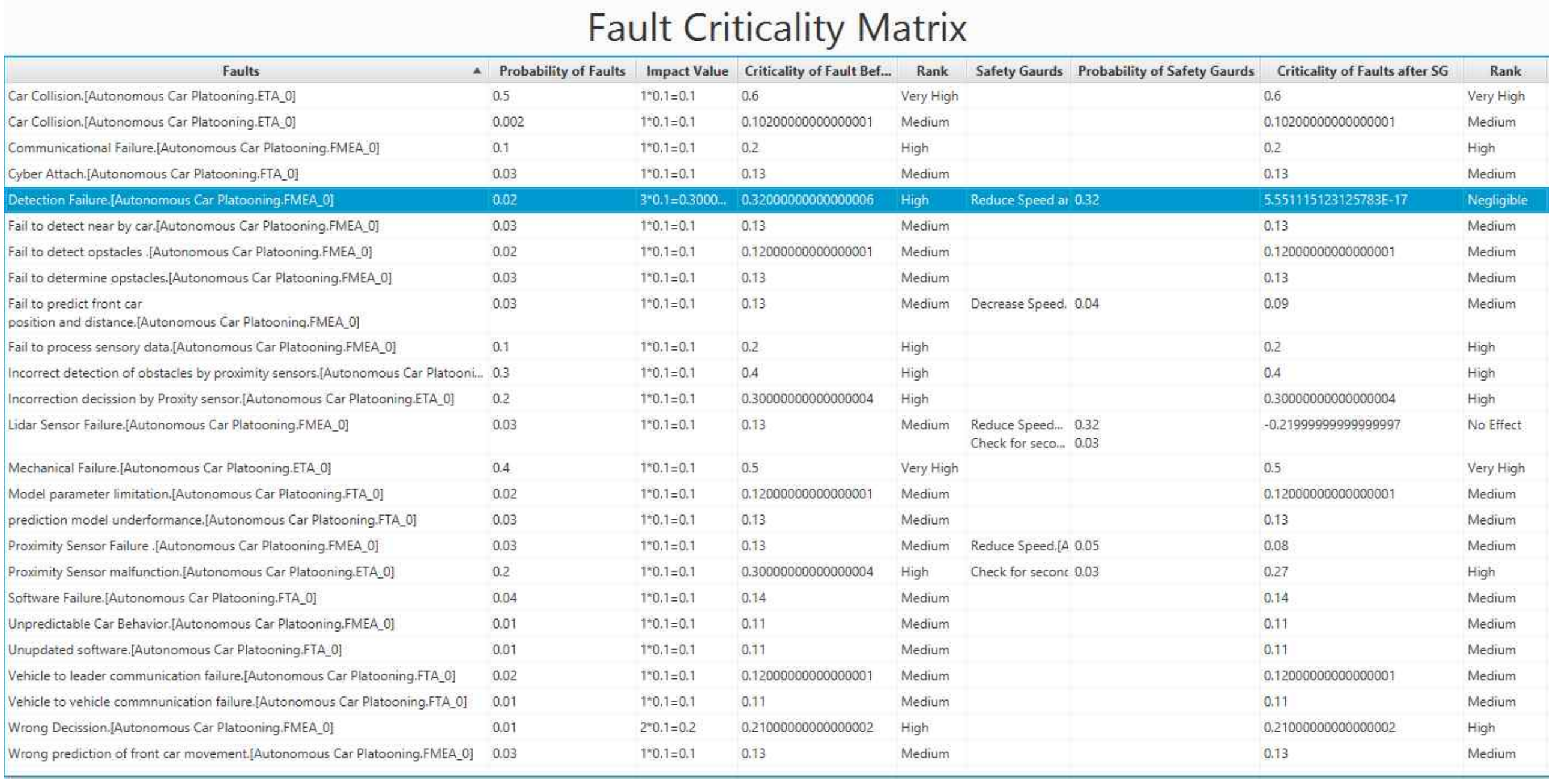
Fault Criticality Matrix

| Faults | Probability of Faults | Impact Value | Criticality of Fault Bef... | Rank | Safety Gaurds | Probability of Safety Gaurds | Criticality of Faults after SG | Rank |
|---|---|---|---|---|---|---|---|---|
| Car Collision.[Autonomous Car Platooning.ETA_0] | 0.5 | 1*0.1=0.1 | 0.6 | Very High | | | 0.6 | Very High |
| Car Collision.[Autonomous Car Platooning.ETA_0] | 0.002 | 1*0.1=0.1 | 0.1020000000000001 | Medium | | | 0.1020000000000001 | Medium |
| Communicational Failure.[Autonomous Car Platooning.FMEA_0] | 0.1 | 1*0.1=0.1 | 0.2 | High | | | 0.2 | High |
| Cyber Attach.[Autonomous Car Platooning.FTA_0] | 0.03 | 1*0.1=0.1 | 0.13 | Medium | | | 0.13 | Medium |
| Detection Failure.[Autonomous Car Platooning.FMEA_0] | 0.02 | 3*0.1=0.3000... | 0.3200000000000006 | High | Reduce Speed ai | 0.32 | 5.551115123125783E-17 | Negligible |
| Fail to detect near by car.[Autonomous Car Platooning.FMEA_0] | 0.03 | 1*0.1=0.1 | 0.13 | Medium | | | 0.13 | Medium |
| Fail to detect opstacles .[Autonomous Car Platooning.FMEA_0] | 0.02 | 1*0.1=0.1 | 0.1200000000000001 | Medium | | | 0.1200000000000001 | Medium |
| Fail to determine opstacles.[Autonomous Car Platooning.FMEA_0] | 0.03 | 1*0.1=0.1 | 0.13 | Medium | | | 0.13 | Medium |
| Fail to predict front car position and distance.[Autonomous Car Platooning.FMEA_0] | 0.03 | 1*0.1=0.1 | 0.13 | Medium | Decrease Speed. | 0.04 | 0.09 | Medium |
| Fail to process sensory data.[Autonomous Car Platooning.FMEA_0] | 0.1 | 1*0.1=0.1 | 0.2 | High | | | 0.2 | High |
| Incorrect detection of obstacles by proximity sensors.[Autonomous Car Platooni... | 0.3 | 1*0.1=0.1 | 0.4 | High | | | 0.4 | High |
| Incorrection decission by Proxity sensor.[Autonomous Car Platooning.ETA_0] | 0.2 | 1*0.1=0.1 | 0.3000000000000004 | High | | | 0.3000000000000004 | High |
| Lidar Sensor Failure.[Autonomous Car Platooning.FMEA_0] | 0.03 | 1*0.1=0.1 | 0.13 | Medium | Reduce Speed...<br>Check for seco... | 0.32<br>0.03 | -0.2199999999999997 | No Effect |
| Mechanical Failure.[Autonomous Car Platooning.ETA_0] | 0.4 | 1*0.1=0.1 | 0.5 | Very High | | | 0.5 | Very High |
| Model parameter limitation.[Autonomous Car Platooning.FTA_0] | 0.02 | 1*0.1=0.1 | 0.1200000000000001 | Medium | | | 0.1200000000000001 | Medium |
| prediction model underformance.[Autonomous Car Platooning.FTA_0] | 0.03 | 1*0.1=0.1 | 0.13 | Medium | | | 0.13 | Medium |
| Proximity Sensor Failure .[Autonomous Car Platooning.FMEA_0] | 0.03 | 1*0.1=0.1 | 0.13 | Medium | Reduce Speed.[A | 0.05 | 0.08 | Medium |
| Proximity Sensor malfunction.[Autonomous Car Platooning.ETA_0] | 0.2 | 1*0.1=0.1 | 0.3000000000000004 | High | Check for seconc | 0.03 | 0.27 | High |
| Software Failure.[Autonomous Car Platooning.FTA_0] | 0.04 | 1*0.1=0.1 | 0.14 | Medium | | | 0.14 | Medium |
| Unpredictable Car Behavior.[Autonomous Car Platooning.FMEA_0] | 0.01 | 1*0.1=0.1 | 0.11 | Medium | | | 0.11 | Medium |
| Unupdated software.[Autonomous Car Platooning.FTA_0] | 0.01 | 1*0.1=0.1 | 0.11 | Medium | | | 0.11 | Medium |
| Vehicle to leader communication failure.[Autonomous Car Platooning.FTA_0] | 0.02 | 1*0.1=0.1 | 0.1200000000000001 | Medium | | | 0.1200000000000001 | Medium |
| Vehicle to vehicle commnunication failure.[Autonomous Car Platooning.FTA_0] | 0.01 | 1*0.1=0.1 | 0.11 | Medium | | | 0.11 | Medium |
| Wrong Decission.[Autonomous Car Platooning.FMEA_0] | 0.01 | 2*0.1=0.2 | 0.2100000000000002 | High | | | 0.2100000000000002 | High |
| Wrong prediction of front car movement.[Autonomous Car Platooning.FMEA_0] | 0.03 | 1*0.1=0.1 | 0.13 | Medium | | | 0.13 | Medium |

Figure 6. Fault Criticality Matrix for the Platooning System.

icality of the fault is reduced to $5.5 * 10^{-17}$ as shown in the column "*Criticality of Fault After SG*". After supplying the safety guard, the criticality rank of the above-mentioned fault becomes "*Negligible*" in the last column that is the "*Rank*" column. This means that the supplied safety guard is enough to mitigate the fault *Detection Failure.[Autonomous Car Platooning System.FMEA_0]*. However, sometimes some faults are so critical that a single safety guard is not enough to mitigate them. They need more safety guards to mitigate or to reduce their criticality from the system. Let us take another example of such a fault that needs more than one safety guard to mitigate its criticality. The probability of fault *Lidar Sensor Failure.[Autonomous Car Platooning.[FMEA_0]]* is 0.03 and it is impacting one other fault due to which its criticality becomes 0.13. As the rank of fault *Lidar Sensor Failure.[Autonomous Car Platooning.[FMEA_0]]* is "*Medium*" before supplying the safety guards. However, a single safety guard was not enough to reduce its criticality, therefore, we supplied two safety guards first one is *Reduce Speed and exit platooning.[Autonomous Car Platooning. [FMEA_0]* and the second one is *Check for the secondary sensor.[Autonomous Car Platooning.[FMEA_0]]*. After supplying these safety guards, we calculated its criticality and the criticality of the fault was reduced to -0.219 and its criticality is now ranked as "*No Effect*" which means the fault has been mitigated and there is no chance of being activated.

However, some faults are very critical that its criticality does not change even after supplying the safety guards. These kinds of faults are dangerous for the system. The criticality of fault *Fails to predict front car position and distance.[Autonomous Car Platooning.[FMEA_0]]* is 0.13 and its rank is "*Medium*" before the safety guard is supplied. However, even after supplying the safety guard *Decrease Speed.[Autonomous Car Platooning.[FMEA_0]]* there was no significant decrease in the criticality of fault is detected as the criticality is still 0.09 which is still huge, and it falls under medium rank, and as we can see that the rank is still "*Medium*" even after the safety guard is supplied. This means that the supplied safety guard is not enough to cope with the fault *Fails to predict front car position and distance.[Autonomous Car Platooning.[FMEA_0]]*. Therefore, safety engineers must give significant importance to such kinds of faults as these faults can be activated in the system at any time.

Some faults in the system do not have safety guards. Therefore, their criticality remains always constant and these kinds of faults are the most critical for system safety. Let us take the example of one such kind of fault and analyze its criticality. The fault *Incorrect decision by the Proximity sensor.[Autonomous Car Platooning.[ETA]]* was discovered by applying ETA and its criticality is 0.3 as there was no safety guard available for this fault. Therefore, both criticality ranking (*i.e., Before and after supplying safety guard.*) is the same which is "*High*". This means that this fault is very critical and can be activated at any time causing the whole system failure as its criticality is quite high. Such kinds of faults must be monitored by a safety engineer as these faults do not have any safety guard and could more safety-critical for CPSs like the platooning system.

### 4.3 Simulation and Evaluation

To evaluate the usefulness of our proposed FCM, we consider the following research questions:

RQ1: How to verify that the criticality of faults provided by the FCM is appropriate? This question is formulated to verify the appropriateness of the criticality of faults in the proposed FCM.

RQ2: How effective is the FCM in ensuring safety in safety-critical systems? This question is formulated to check whether proposed safety guards in FCM reduce the risk associated with the faults.

In response to the RQ1, we refer to the guidelines from ISO 26262 standard to verify the appropriateness of the faults and their criticality in our FCM. This is to check that the level of faults criticality in our approach is at a similar level compared to the ISO 26262 standard and [16]. ISO 26262 standard widely used in automotive industries to ensure safety. It defines the three safety attributes such as *severity, exposure, and controllability* to construct criticality levels of faults in safety-critical systems [15]. The authors [16] used the guidelines of ISO 26262 to perform the hazard analysis of the heavy-truck platooning. They used the ranking system of ISO 26262 to rank the criticality of the faults based on their severity, exposure, and controllability as shown in Table 3.

In ISO 26262, the *severity* refers to the extent of harm to an individual in a specific situation. The severity has four levels as mentioned below:

- S0: No injuries; S1: Light to moderate injuries; S2: Severe to life-threatening injuries; S3: Life-threatening to fatal injuries.

*Exposure* refers to the relative frequency of the operational conditions, in which hazardous events may occur

Table 3. Example of faults considered in hazard analysis and risk assessment of heavy-truck platooning in [16].

| Faults | Accidents | Severity | Exposure | Controllability |
|---|---|---|---|---|
| Communication failure | Crash | S3 | E2 | C2 |
| Onboard sensors failure | Platooning system error | S2 | E1 | C3 |
| Stopped vehicle on the road | Collision with vehicle | S3 | E3 | C2 |
| Detection failure | Crash | S3 | E1 | C2 |
| Weather (fog, rain, sleet) | Operation in a dangerous condition | S3 | E2 | C1 |

and cause hazards and injuries. Its five levels are as below;

- E0: incredibly unlikely; E1: Very low probability of exposure; E2: Low probability of exposure; E3: Medium probability of exposure; E4: High probability of exposure.

*Controllability* refers to the prevention of specific damage through the timely reactions of the persons. The levels of controllability are given as follows.

- C0: Controllable in general; C1: Easily controllable; C2: Normally controllable; C3: Uncontrollable

The faults in our FCM to perform criticality analysis of the platooning were carefully considered after following the guidelines from ISO 26262 and other scientific literature such as [14], and [16]. We considered the following faults from the platooning case study to see that the criticality ranking of faults in our FCM is at a similar level compared to the criticality ranking of faults in [16].

- *Communication failure, cyber-attack, detection failure, fail to detect a nearby car, fail to detect obstacles, fail to predict front car position and distance, fail to process sensory data, LIDAR sensor failure, mechanical failure, proximity sensor failure, and software failure*, etc.

As shown in Figure 6. the FCM shows the criticality of faults before applying safety guards and after applying safety guards. We see that the criticality of faults in Figure 6 are at a similar level compared to the criticality of faults in [16]. Although, we have ranked the faults into seven categories in the platoon driving system as shown in Table 2. However, the fault ranking falls under the general ranking of faults in ISO 26262. The reason to rank the faults in more categories in our FCM is to define more concrete controllability mechanisms according to the criticality value of faults.

The description of the faults in our FCM (Figure 6) and the description of faults in existing literature may differ to some extent. However, the types and the class of faults are the same. All faults considered in the FCM are well-known faults associated with the platooning system. For example, the description of the fault "*detection failure*" in our FCM and the description of the fault "*onboard sensor failure*" in [16] is different. However, the type of fault and class of fault is the same and both faults belong to the detection system type with high criticality.

In order to verify the reduction of risk as seen in our FCM (Figure 6), we used VEhicular NeTwork Open Simulator (VENTOS) [17] to analyze safety for the platooning system. The VENTOS is a closed-loop VANET simulator that combines the capabilities of both communication networks and vehicular traffic simulators. It is an open-sourced simulator that comprises Simulation of Urban Mobility (SUMO) and Objective Modular Network Testbed in C++ (OMNET++). It was designed for vehicular traffic follow analysis and can incorporate new control logic, such as intelligent traffic controller, collaborative driving (e.g., the platooning driving) dynamic routing, and autonomous driving capability.

We choose this simulator to verify the safe behavior of the platooning driving because it supports the platooning driving system. In our simulation setup, we first make a platoon of six vehicles (V1, V2....., V6) as shown in Figure 7. The red-colored car is the leader of the platoon and blue color cars are followers in the platoon. We first present a scenario in which the Platooning system was not analyzed using FCM.

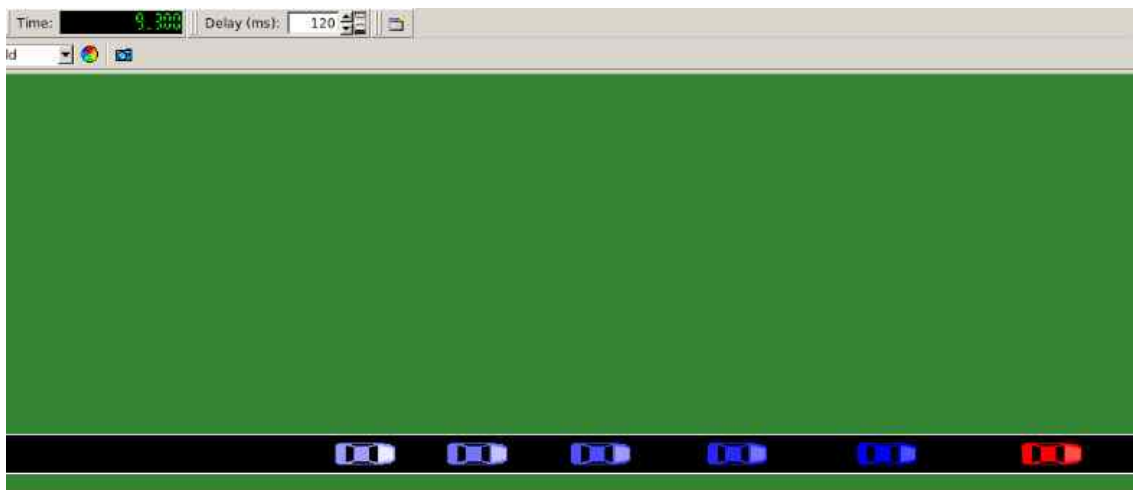

Figure 7. A Platoon of Six Vehicles.

Scenario 1: The platoon was moving towards its destination under normal conditions. However, at some point,

suddenly the leader faces a non-platooning car (i.e., the yellow car). The leader sends commands to the follower vehicles to reduce the speed. However, due to the narrow distance between the vehicle in the platoon and less reaction time to react the leader's command which results in a collision of the platoon. Figure 8 shows the collision of the platoon in the VENTOS simulator.

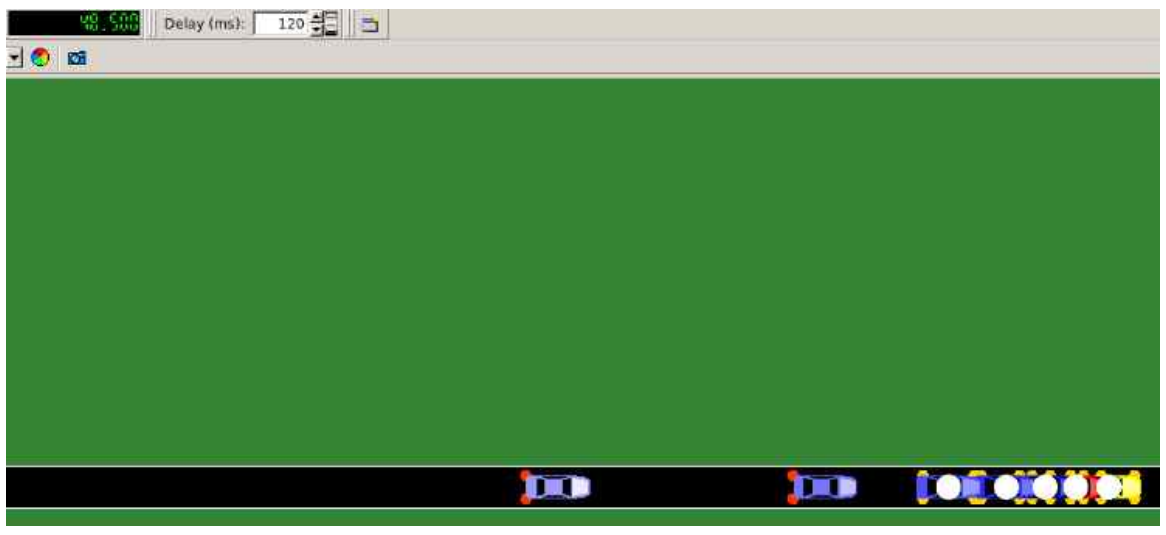

Figure 8. Platooning Collision without Analyzing Criticality of Faults using FCM

Scenario 2: In the second scenario, we analyze the criticality of potential faults and make arrangements for safety guards if any critical fault is triggered among vehicles in the platoon. Hence, we perform the criticality analysis of faults using our FCM as shown in Figure 6. This time we simulate a scenario in which we first perform the criticality analysis of faults of the platooning system then we simulate a platoon of 8 vehicles in the VENTOS simulator. The platoon was driving towards its destination under normal conditions. At some point, the leader received a message from Road Side Unit (RSU) about a traffic accident on the lane on which the platoon is driving. The leader transmits a command to activate safety guards such as "*reduce the speed*" and "*change the lane*". Hence, the followers receive the command from the leader and make safe lane change by reducing their speed. Figure 9 shows the safe behavior of the platooning after activating safety guards discovered during fault criticality analysis.

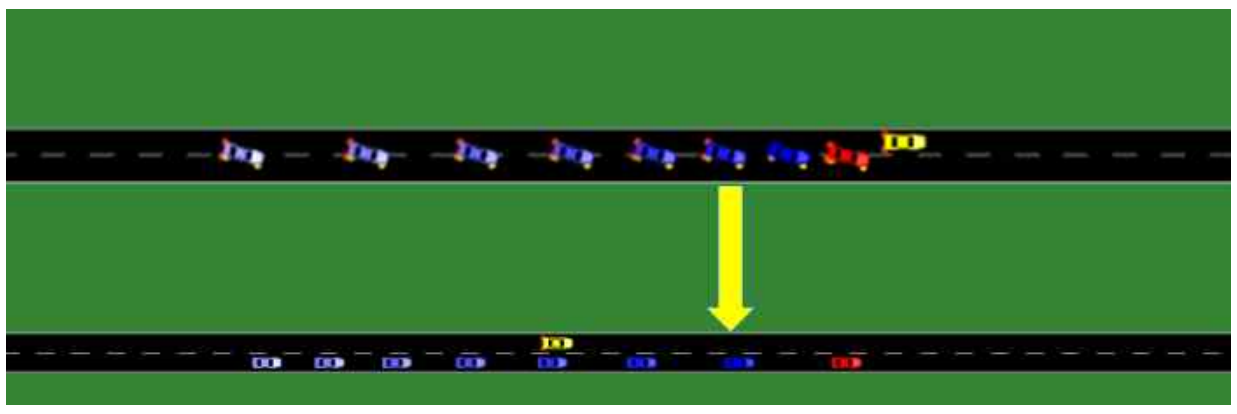

Figure. 9. Activation of the Safety guards (i.e., reduction in speed and safe lane change) in the Platooning driving.

In response to the research question RQ2, Figures 6, 8, and 9 show the effectiveness of the FCM. As compared to the scenario in which the platooning driving's fault criticality was analyzed using FCM, it is safer than the scenario in which the fault criticality of the platooning driving was not analyzed with FCM.

It is cleared from the presented scenario 1 in which the platooning system was not analyzed using FCM faces a crash due to the "*detection failure*". In this scenario platoon leader suddenly faces a non-platooning vehicle, due to less reaction time the leader could not apply an emergency brake to stop the vehicle. Therefore, the platooning system crashed. Figure 8 shows the hazardous scenario of the platooning system without analyzing the criticality of faults using the FCM.

In scenario 2 the platooning system was analyzed by the FCM. Hence, all the potential faults were considered during the criticality analysis and arranged safety guards to cope with the faults in the future. In this scenario, the leader received a message from RSU about a fault i.e., "*traffic accident*" on the lane on which the platoon is driving. Hence the platoon leader activates the predefined safety guards (i.e., *decreasing speed and safe lane change*) and sends the command to activate the safety guards for followers. The followers receive the command from the leader and made a safe lane change by decreasing the speed. Hence, the safety of the platooning in the hazardous scenario is ensured by activating the timely mannered safety guards. Figure 9 shows the good impact of the FCM on the safety of the platooning in which the platoon system's safety was ensured by activating safety guards (i.e., *decreasing speed and safe lane change*).

To verify the high critical fault's impact on the safety of the platoon system, we present another scenario in which the platooning system faces a fault i.e., *Communication Failure*. The criticality of this fault is *0.2* which is ranked as "*High*" in the rank column in Figure 6. As we can see that the severity of the *communication failure* is *S2* in the hazard analysis and risk assessment table (i.e., Table 3.) of the platooning system in [16]. The severity of this fault falls under the "*Severe to life-threatening injuries*" rank according to ISO 26262 [15]. Due to *communication failure* between the follower and leader vehicle, the follower vehicle made sudden acceleration, and hence the platoon crash has happened. As we can see the rank of the fault *communication failure* is high and the chance of crash was also higher, hence during the simulation the crash has happened which shows the correct-ness and effectiveness of calculated criticality of the fault *communication failure* in the FCM.

In the last scenario, we present another important potential fault i.e., *detection failure.* In this scenario, as the *detection failure* rank is *high* in Figure 6, so there is a huge chance of the platooning collision. The severity of the fault *detection failure* is *S3* in Table 3 of this hazard analysis and risk assessment of the platooning system in [16]. This fault's ranking falls under the "*Life-threatening to fatal injuries*" rank according to ISO 26262. Such kinds of faults are very dangerous to the safety of the vehicle as well as for the safety of human lives. Therefore, we arranged three safety guards for this fault. The platooning was driving in normal condition, at some point, the system detects the decrease in vision system of the leader vehicle due to environmental condition (i.e., dense fog). The system detects high critical faults i.e., *detection failure,* and activates the safety guards. The platooning leader sends the commands to followers to activate safety guards such as *reduce speed, activate ACC mode, and dissolve platooning*. Hence the safety of each vehicle in the platooning increased by activating safety guards.

### 4.4 Threats to Validity

Internal Validity: Our proposed FCM is based on the four content relationships (influence, inheritance, overlap, and supplement relationship) among the hazard analysis (i.e., FTA, FMEA, and ETA) artifacts only. The first threat to internal validity is that we consider the above-mentioned four types of relationships based on the FTA, FMEA, and ETA in the composite hazard analysis of CCPS. There exists other hazards analysis techniques like STPA (System Theoretic Process Analysis) and etc. which are not considered in our proposed approach. However, FTA, ETA and FMEA are most commonly used hazard analysis techniques to ensure safety for safety-critical systems.

External Validity: We used the VENTOS simulator to verify the effectiveness of FCM in our case study. VENTOS simulator is only dedicated to the Platooning CPS. Therefore, the result may differ based on the domain of the CCPS and selected simulator according to the chosen domain. For example, if we analyze the criticality of faults associated with a smart manufacturing system using FCM, then we need to use a simulator related to smart manufacturing CPS. Therefore, this technique poses a threat in terms of the generalizability of the verification part of the approach. However, we minimize this threat by making the FCM part more generalize which is our main contribution. Any collaborative CPS's fault criticality can be analyzed by using our FCM. However, for particular CCPS, we need a separate simulator in order to verify the effectiveness of FCM.

## 5. Conclusion and Future Work

Collaborative Cyber-Physical Systems (CCPS) are those systems that contain tightly coupled physical and cyber components, massively interconnected subsystems, jointly collaborating and exchanging information and resources, and performing an individual task to achieve a common goal. The failure in one system in collaborative CPSs may lead to activating many other faults in participant systems. Safety assurance in CCPS is a thorny challenge for safety engineers and the determination of criticality of faults is also a challenging task especially in complex systems like collaborative CPSs.

Therefore, we proposed the Fault Criticality Matrix for fault criticality analysis based on the composite hazard analysis technique. This matrix presents the criticality of faults with an impact value of the fault. We analyze the hazards for the platooning system with our composite hazard analysis tool and perform the criticality analysis using FCM. After performing the repeated fault criticality analysis of the platooning system, our simulation results show that safety was assured.

As our project work is going on and in the future, we will work to develop a machine-learning algorithm to provide sustainable and resilient safety for CPSs in on-the-fly situations.

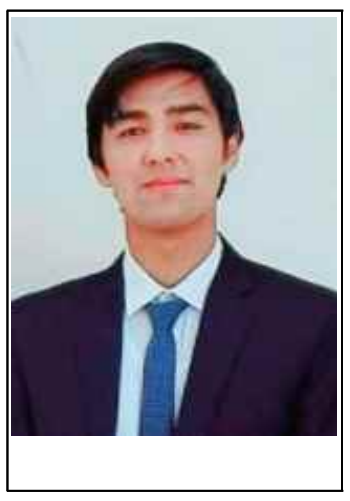

## Manzoor Hussain

https://orcid.org/0000-0002-9131-0930
e-mail : hussain@selab.cbnu.ac.kr

He received Bachelor degree at the Department of Software Engineering, The Univeristy of Azad Jammu and Kashmir Muzaffarabard, Pakistan in 2017 Currently, he is a graduate student in the Department of Computer Science Chungbuk National University, Korea sinece 2020. His research interests are in the area of Software Engineering, Deep Learning, Autonomous Car, and Cyber Physical Systems

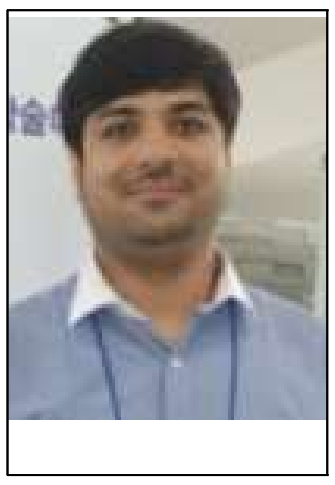

## Nazakat Ali

http://orcid.org/0000-0002-3875-812X
e-mail : nazakatali@selab.cbnu.ac.kr

He received a Master degree and a Ph.D. degree at the Department of Computer Science, Chungbuk National University, Korea, in 2018 and 2021, repectively. Currently, he is working as a Postdoc in Software Engineering Laboratory, Department of Computer Science, Chungbuk National University. His research interests are in the area of Software requirements engineering, Ontology, software architecture, DevOps, and Safety for Cyber-Physical Systems

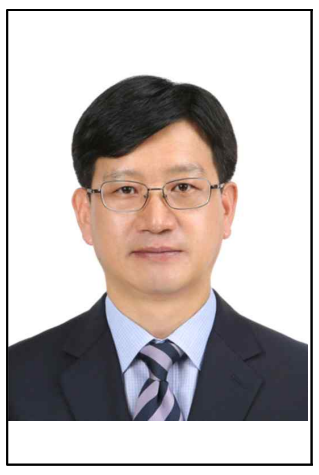

## Jang-Eui Hong

https://orcid.org/0000-0001-9786-7732
e-mail : jehong@chungbuk.ac.kr

He receivec a Ph.D degree at the Dapartment of Computer Science KAIST, 2001. He served as a researcher in Agency of Defense Development, Korea, and also worked for Solution-Link Co. as a Technical Director. Currently he is a professor in the Department of Computer Science, Chungbuk National University, Korea. His research interests are in the areas of Model-Based Software Engineering. Software Quality, Cyber-Physical Systems.